\documentclass[twocolumn]{aastex62}
\usepackage{amssymb}
\usepackage{natbib}
\usepackage{hyperref}
\usepackage{color}
\usepackage{pdfcomment}

\newcommand\changed[1]{{#1}}

\received{May 2, 2018}
\revised{May 21, 2018}
\accepted{May 23, 2018}
\submitjournal{AJ}

\shorttitle{The Reactivation and Nucleus Characterization of 358P/PANSTARRS}
\shortauthors{Hsieh et al.}

\begin{document}

\title{The Reactivation and Nucleus Characterization of Main-Belt Comet 358P/PANSTARRS (P/2012 T1)
}

\correspondingauthor{Henry H.\ Hsieh}
\email{hhsieh@psi.edu}

\author[0000-0001-7225-9271]{Henry H.\ Hsieh}
\affil{Planetary Science Institute, 1700 East Fort Lowell Rd., Suite 106, Tucson, AZ 85719, USA}
\affil{Institute of Astronomy and Astrophysics, Academia Sinica, P.O.\ Box 23-141, Taipei 10617, Taiwan}

\author[0000-0002-7332-2479]{Masateru Ishiguro}
\affil{Department of Physics and Astronomy, Seoul National University, Gwanak, Seoul 151-742, Korea}


\author[0000-0003-2781-6897]{Matthew M.\ Knight}
\affil{Department of Astronomy, University of Maryland, 1113 Physical Sciences Complex, Building 415, College Park, MD 20742, USA}


\author[0000-0001-7895-8209]{Marco Micheli}
\affil{ESA SSA-NEO Coordination Centre, Largo Galileo Galilei, 1, 00044 Frascati (RM), Italy}
\affil{INAF - Osservatorio Astronomico di Roma, Via Frascati, 33, 00040 Monte Porzio Catone (RM), Italy}

\author[0000-0001-6765-6336]{Nicholas A.\ Moskovitz}
\affil{Lowell Observatory, 1400 W.\ Mars Hill Rd, Flagstaff, AZ 86001, USA}

\author[0000-0003-3145-8682]{Scott S.\ Sheppard}
\affil{Department of Terrestrial Magnetism, Carnegie Institution for Science, 5241 Broad Branch Road NW, Washington, DC 20015, USA}


\author[0000-0001-9859-0894]{Chadwick A.\ Trujillo}
\affil{Department of Physics and Astronomy, Northern Arizona University, Flagstaff, AZ 86001, USA}



\begin{abstract} 
We present observations of main-belt comet 358P/PANSTARRS (P/2012 T1) obtained using the Gemini South telescope from 2017 July to 2017 December, as the object approached perihelion for the first time since its discovery.  We find best-fit IAU phase function parameters of \changed{$H_R=19.5\pm0.2$~mag and $G_R=-0.22\pm0.13$} for the nucleus, corresponding to an effective radius of \changed{$r_N=0.32\pm0.03$~km} (assuming an albedo of $p_R=0.05$).  The object appears significantly brighter (by $\geq1$~mag) than expected starting in 2017 November, while a faint dust tail oriented approximately in the antisolar direction is also observed on 2017 December 18.  We conclude that 358P has become active again for the first time since its previously observed active period in 2012-2013.  These observations make 358P the seventh main-belt comet candidate confirmed to exhibit recurrent activity near perihelion with intervening inactivity away from perihelion, strongly indicating that its activity is sublimation-driven.  Fitting a linear function to the ejected dust masses inferred for 358P in 2017 when it is apparently active, we find an average net dust production rate of \changed{$\dot M = 2.0\pm0.6$~kg~s$^{-1}$} (assuming a mean effective particle radius of $\bar a_d=1$~mm) and an estimated activity start date of 2017 November $8\pm4$ when the object was at a true anomaly of $\nu=316^{\circ}\pm1^{\circ}$ and a heliocentric distance of $R=2.54$~AU.  Insufficient data is currently available to ascertain whether activity strength has changed between the object's  2012-2013 and 2017 active periods.  Further observations are therefore highly encouraged during the object's upcoming observing window (2018 August through 2019 May).
\end{abstract}




\hyphenation{re-arr-ang-ing}

\section{INTRODUCTION}\label{section:introduction}
\subsection{Background}\label{section:background}

Comet 358P/PANSTARRS, formerly designated P/2012 T1 (PANSTARRS), was discovered on UT 2012 October 6 by the Pan-STARRS1 (PS1) survey telescope on Haleakala \citep{wainscoat2012_p2012t1}.  At the time of its discovery, it was the seventh object to be identified as a likely main-belt comet (MBC) \citep{hsieh2013_p2012t1,moreno2013_p2012t1}, where about a dozen such objects are known to date.

MBCs exhibit activity in the form of comet-like dust emission that has been determined to be at least partially due to the sublimation of volatile ice, yet occupy stable orbits in the main asteroid belt \citep{hsieh2006_mbcs}.  \changed{They comprise a subset of the population of active asteroids \citep{jewitt2015_actvasts_ast4}, which include all objects that exhibit comet-like activity due to a variety of mechanisms or combination of mechanisms, including sublimation, impact disruption, and rotational destabilization,  yet have dynamically asteroidal orbits.  Small solar system bodies are commonly considered dynamically asteroidal if they have Tisserand parameter values of $T_J>3$ \citep{kresak1979_cometasteroidinterrelations_ast1}, although in practice, the dynamical transition zone between asteroids and comets actually appears to lie roughly between $T_J=3.05$ and $T_J=3.10$ \citep{tancredi2014_asteroidcometclassification,jewitt2015_actvasts_ast4,hsieh2016_tisserand}.}

While some MBCs may be Jupiter-family comets (JFCs) that have recently evolved onto main-belt-like orbits, others reside in regions of orbital element space that are largely unreachable by interloping JFCs, and could potentially have formed in situ \citep{hsieh2016_tisserand}.  This raises the intriguing possibility that they might be able to help constrain the temperature and composition of objects in this region in the early solar system, and also provide a means to explore the possibility that icy material originally from the main-belt region of the solar system, or at least icy material similar in composition to objects currently occupying the main asteroid belt, could have been a significant primordial source of terrestrial water \citep[][and references within]{morbidelli2000_earthwater,raymond2004_earthwater,raymond2017_waterorigin,obrien2006_earthwater,obrien2018_waterdelivery}.

358P is one of four MBCs that were found to have orbital elements (specifically relatively high eccentricities and inclinations) similar to those taken on by test particles that were found by \citet{hsieh2016_tisserand} to temporarily transition from initially JFC-like orbits to main-belt-like orbits in numerical integrations.  In that work, however, most test particles exhibiting that type of dynamical behavior were not found to remain on their adopted main-belt orbits for very long ($<$20~Myr). Numerical integrations specifically focused on 358P found that the object is largely dynamically stable over a timescale of 100~Myr \citep{hsieh2013_p2012t1}, suggesting that the likelihood of a JFC-like origin, while non-zero, is low.

The direct detection of sublimation products (i.e., gas or vapor) from MBCs has proven to be extremely challenging to achieve using currently available observational facilities \citep{snodgrass2017_mbcs}. Attempts to specifically search for outgassing from 358P using the Keck Observatory, Very Large Telescope, and {\it Herschel Space Telescope} yielded only upper limit production rates of $Q({\rm CN})<1.5\times10^{23}$~molecules~s$^{-1}$, $Q({\rm H_2O})<7.6\times10^{25}$~molecules~s$^{-1}$, and $Q({\rm OH})<6\times10^{25}$~molecules~s$^{-1}$, respectively, all of which roughly correspond to similar upper limit water production rates of $\sim8\times10^{25}$~molecules~s$^{-1}$ \citep{hsieh2013_p2012t1,orourke2013_p2012t1,snodgrass2017_358p}.  As such, for now, the identification of the driver of an active asteroid's activity must be accomplished by indirect methods such as numerical dust modeling or the identification of recurrent activity after intervening periods of inactivity.

Dust modeling can allow for the determination of the duration of the period of active dust emission for an object, which can then indicate what mechanism or combination of mechanisms is likely to be responsible for that emission event. A short-duration (i.e., impulsive) dust emission event is most likely to be caused by an impact, while sublimation and rotational disruption are more plausible explanations for longer lasting dust emission events.  In addition to ambiguity as to whether a long-lasting emission event could have been caused by sublimation or rotational disruption, though, dust modeling is also susceptible to parameter degeneracies which could lead to misleading results depending on parameter spaces and emission scenarios being explored \citep[cf.][]{hsieh2012_scheila}.

Meanwhile, recurrent activity interspersed with periods of inactivity, particularly if it is periodic and occurs near perihelion, is considered to be an extremely strong indication that an object's activity is sublimation-driven, given that such behavior cannot be plausibly explained by any other proposed mechanism \citep[e.g.,][]{jewitt2015_actvasts_ast4}.  As such, the search for and identification of recurrent activity for MBC candidates has been a key component of the study of active asteroids over the last several years \citep[e.g.,][]{hsieh2010_133p,hsieh2014_176p,hsieh2016_238p,hsieh2015_324p,agarwal2016_288p_cbet}.  Identification of activity is often achieved by simple visual detection of an extended coma or tail, or analysis of an object's point-spread function (PSF) compared to those of nearby background stars. If the precise absolute magnitude (as well as, ideally, basic lightcurve properties) of an object's inactive nucleus is known, however, much more sensitive searches can be conducted via searches for photometric enhancements indicative of unresolved dust emission \citep[e.g.,][]{tholen1988_chiron,hartmann1989_chiron,hsieh2015_324p}.

Knowledge of the absolute magnitude of an object's nucleus furthermore facilitates detailed analyses of activity strength during active periods by enabling the precise determination of the flux contribution from emitted dust via the subtraction of the nucleus contribution from the total flux measured for a MBC while it is active.  As such, physical characterization of MBC nuclei, while also being valuable for improving our understanding of the physical characteristics of the overall MBC population and providing constraints for thermal models, also comprises an important complement to activity searches and characterization efforts.

In this work, we present observations obtained to physically characterize the nucleus of MBC 358P/PAN-STARRS, as well as report on the discovery of its reactivation in late 2017.

\section{Observations}\label{section:observations}

\begin{figure}[ht!]
\centerline{\includegraphics[width=\columnwidth]{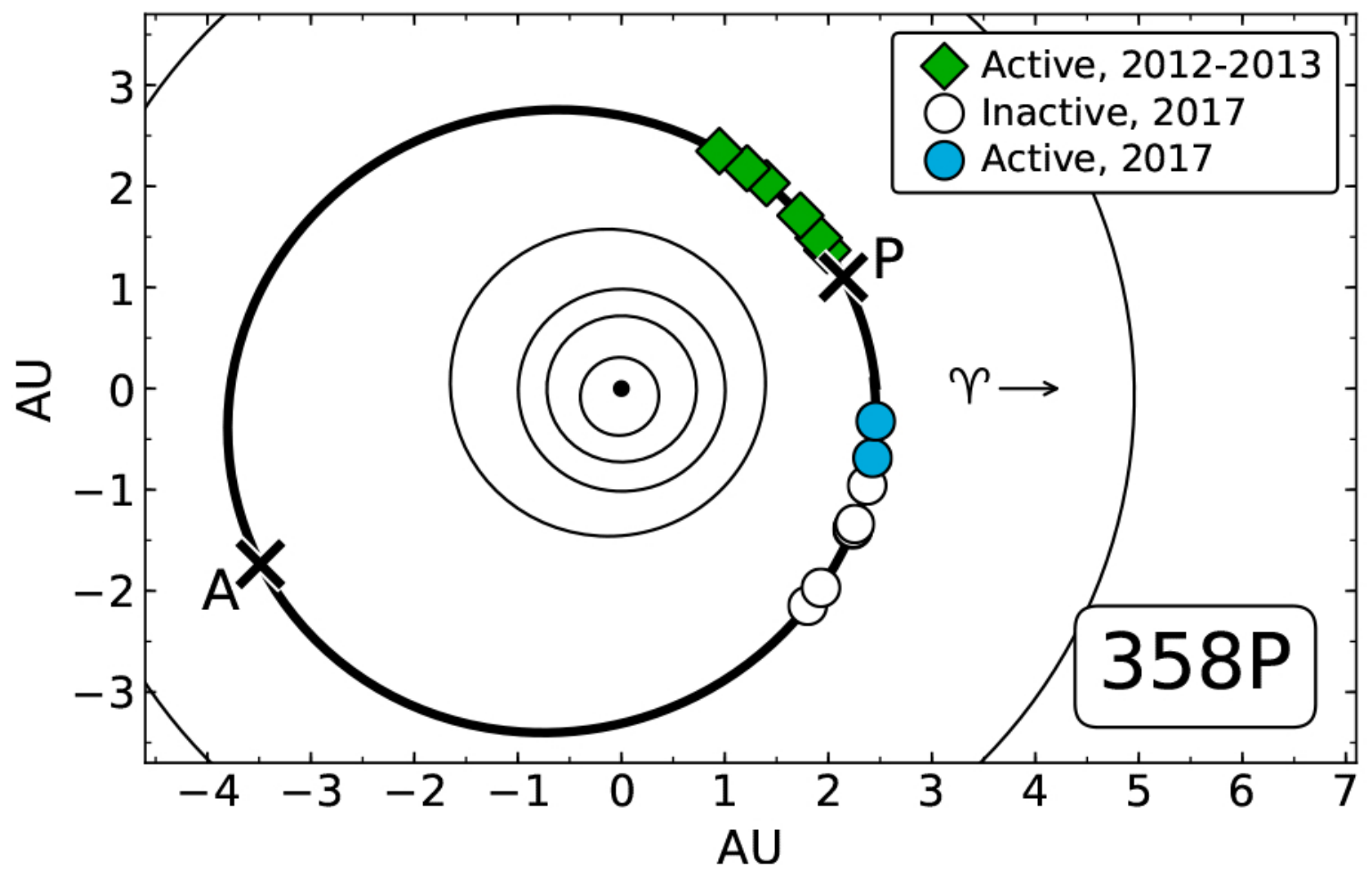}}
\caption{\small Orbit position plot with the Sun (black dot) at the center, and the orbits of Mercury, Venus, Earth, Mars, 358P, and Jupiter shown as black lines. Perihelion (P) and aphelion (A) are marked with crosses. Green diamonds mark positions of observations when 358P was active in 2012-2013, open circles mark positions of observations when 358P was apparently inactive in 2017, and blue circles mark positions of observations when 358P was active in 2017.
}
\label{figure:orbit_358p}
\end{figure}

\begin{figure*}[htb!]
\centerline{\includegraphics[width=5in]{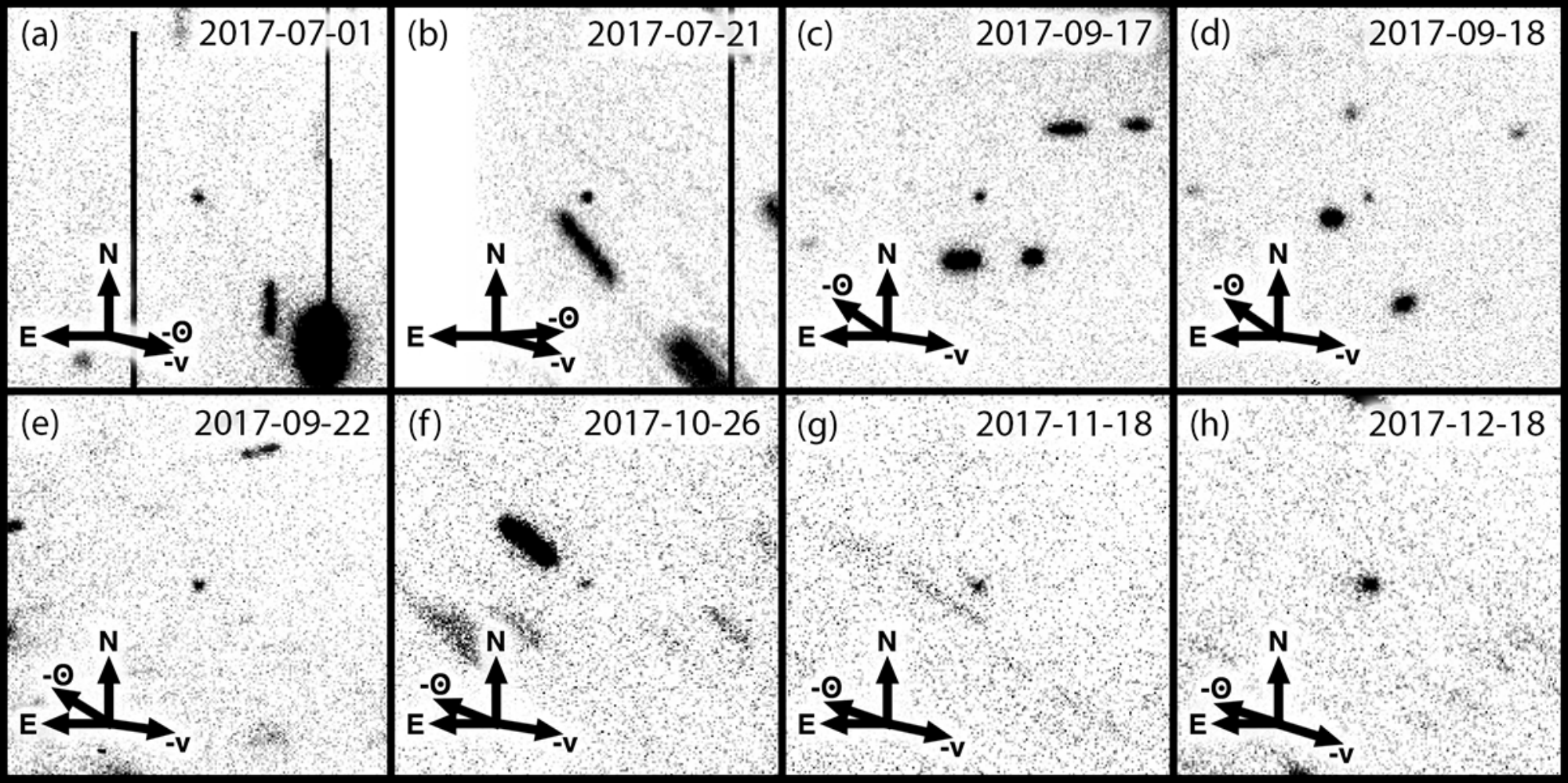}}
\caption{\small Composite $r'$-band images of 358P (at the center of each panel) constructed from data listed in Table~\ref{table:obs_358p}, where no evidence for activity is found for the object at the time of the observations in panels (a) through (f), while we have determined that the object is likely to be active at the time of the observations in panels (g) and (h).  All panels are $30''\times30''$ in size, with north (N), east (E), the antisolar direction ($-\odot$), and the negative heliocentric velocity vector ($-v$), as projected on the sky, marked.  Observation dates (in YYYY-MM-DD format) are listed in the upper right corner of each panel.
}
\label{figure:images_358p}
\end{figure*}

Observations of 358P presented here were obtained with the 8.1~m Gemini South (Gemini-S) telescope at Cerro Pachon (Gemini Program IDs GS-2017A-LP-11 and GS-2017B-LP-11).  We employed the Gemini Multi-Object Spectrograph \citep[GMOS;][]{hook2004_gmos,gimeno2016_gmoss} in imaging mode, a Sloan $r'$-band filter, non-sidereal tracking, and 300~s individual exposure times for all observations.  All observations were conducted at airmasses of $<1.8$, and random dither offsets of up to $10''$ east or west, and north or south were applied to each individual exposure.  Standard bias subtraction, flatfield correction, and cosmic ray removal were performed for all images using Python 3 code utilizing the {\tt ccdproc} package in Astropy\footnote{\url{http://www.astropy.org}} \citep{astropy2018_astropy} and the {\tt L.A.Cosmic} python module\footnote{Written for python by Maltes Tewes (\url{https://obswww.unige.ch/\~tewes/cosmics\_dot\_py/})} \citep{vandokkum2001_lacosmic}.

Photometry measurements of the target object and at least five background reference stars were performed using IRAF software \citep{tody1986_iraf,tody1993_iraf}, with absolute photometric calibration performed using field star magnitudes from the PS1 field star catalogs \citep{schlafly2012_ps1,tonry2012_ps1,magnier2013_ps1}.  Conversion of $r'$-band Gemini and PS1 photometry to $R$-band was accomplished using transformations derived by \citep{tonry2012_ps1} and by R.\ Lupton ({\tt http://www.sdss.org/}).  Target photometry was performed using circular apertures with sizes chosen using curve-of-growth analyses of each night of data, where background statistics were measured in nearby but non-adjacent regions of blank sky to avoid potential dust contamination from the object or nearby field stars.  To maximize signal-to-noise ratios (S/N), we construct composite images of the object for each night of data by shifting and aligning individual images on the object's photocenter using linear interpolation and then adding them together.

Details of our observations of 358P are listed in Table~\ref{table:obs_358p}, where we also mark the orbit positions of both the observations reported here and observations previously reported in \citet{hsieh2013_p2012t1} in Figure~\ref{figure:orbit_358p}.  Composite images of the object during each night of observations reported in this work are shown in Figure~\ref{figure:images_358p}.

\setlength{\tabcolsep}{4.5pt}
\setlength{\extrarowheight}{0em}
\begin{table*}[htb!]
\caption{358P Observations}
\centering
\smallskip
\footnotesize
\begin{tabular}{lcrrccrccrccc}
\hline\hline
\multicolumn{1}{c}{UT Date}
 & \multicolumn{1}{c}{Tel.$^a$}
 & \multicolumn{1}{c}{$N$$^b$}
 & \multicolumn{1}{c}{$t$$^c$}
 & \multicolumn{1}{c}{Airmass$^d$}
 & \multicolumn{1}{c}{Filter}
 & \multicolumn{1}{c}{$\nu$$^e$}
 & \multicolumn{1}{c}{$R$$^f$}
 & \multicolumn{1}{c}{$\Delta$$^g$}
 & \multicolumn{1}{c}{$\alpha$$^h$}
 & \multicolumn{1}{c}{$m_{R,n}$$^i$}
 & \multicolumn{1}{c}{$m_{R,n}(1,1,\alpha)$$^j$}
 & \multicolumn{1}{c}{Active?$^k$}
 \\
\hline
2017 Jul 01 & Gemini-S &  3 &   900 & 1.03 & $r'$ & 284.9 & 2.801 & 2.005 & 15.4 & 24.7$\pm$0.1 & 20.8$\pm$0.1 & no  \\ 
2017 Jul 21 & Gemini-S &  3 &   900 & 1.04 & $r'$ & 289.3 & 2.756 & 1.816 & 10.0 & 24.0$\pm$0.1 & 20.5$\pm$0.1 & no  \\ 
2017 Sep 17 & Gemini-S &  2 &   600 & 1.00 & $r'$ & 302.9 & 2.633 & 1.805 & 15.0 & 24.3$\pm$0.1 & 20.9$\pm$0.1 & no  \\ 
2017 Sep 18 & Gemini-S &  1 &   300 & 1.00 & $r'$ & 303.1 & 2.631 & 1.811 & 15.3 & 24.4$\pm$0.2 & 21.0$\pm$0.2 & no  \\ 
2017 Sep 22 & Gemini-S &  3 &   900 & 1.10 & $r'$ & 304.1 & 2.623 & 1.839 & 16.4 & 24.2$\pm$0.1 & 20.8$\pm$0.1 & no  \\ 
2017 Oct 26 & Gemini-S &  1 &   300 & 1.01 & $r'$ & 312.7 & 2.560 & 2.154 & 22.3 & 25.0$\pm$0.4 & 21.3$\pm$0.4 & no  \\ 
2017 Nov 18 & Gemini-S &  3 &   900 & 1.22 & $r'$ & 318.8 & 2.522 & 2.404 & 23.0 & 24.2$\pm$0.1 & 20.3$\pm$0.1 & yes \\ 
2017 Dec 18 & Gemini-S &  5 &  1500 & 1.67 & $r'$ & 327.0 & 2.479 & 2.722 & 21.1 & 23.2$\pm$0.1 & 19.1$\pm$0.1 & yes \\ 
\hline
\hline
\multicolumn{13}{l}{$^a$ Telescope used.} \\
\multicolumn{13}{l}{$^b$ Number of exposures.} \\
\multicolumn{13}{l}{$^c$ Total integration time, in seconds.} \\
\multicolumn{13}{l}{$^d$ Average airmass of observations.} \\
\multicolumn{13}{l}{$^e$ True anomaly, in degrees.} \\
\multicolumn{13}{l}{$^f$ Heliocentric distance, in AU.} \\
\multicolumn{13}{l}{$^g$ Geocentric distance, in AU.} \\
\multicolumn{13}{l}{$^h$ Solar phase angle (Sun-object-Earth), in degrees.} \\
\multicolumn{13}{l}{$^i$ Equivalent mean apparent $R$-band nucleus magnitude.} \\
\multicolumn{13}{l}{$^j$ Equivalent mean reduced $R$-band nucleus magnitude.} \\
\multicolumn{13}{l}{$^k$ Is activity detected?} \\
\end{tabular}
\label{table:obs_358p}
\end{table*}

\section{Results and Analysis}

\subsection{Phase Function Determination}\label{subsection:phase_function}

\begin{figure}[ht!]
\centerline{\includegraphics[width=\columnwidth]{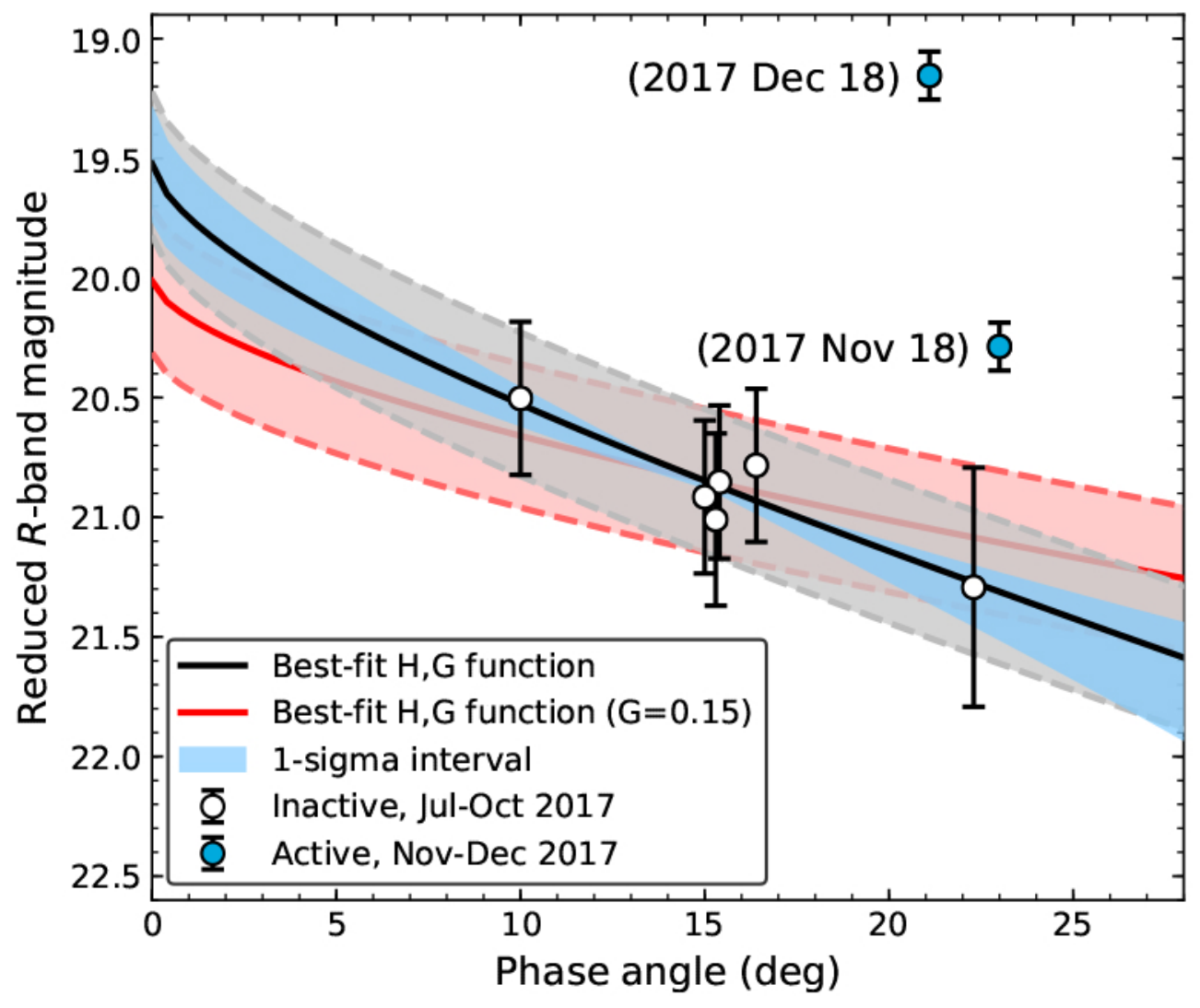}}
\caption{\small Plot of best-fit IAU phase functions for 358P with photometric points measured in 2017 overplotted \changed{(with plotted uncertainties incorporating both measured photometric uncertainties and estimated uncertainties due to the unknown rotational phases at which observations were obtained)}, where the best-fit IAU phase function with $H_R$ and $G_R$ as free parameters is marked with a solid black line and the best-fit IAU phase function assuming $G_R=0.15$ is marked with a solid red line.  The blue-shaded region indicates the 1-sigma range of uncertainty due to phase function parameter uncertainties for the best-fit phase function with $G_R$ as a free parameter, while the gray-shaded and light-red-shaded regions (bounded by gray dashed lines and red dashed lines, respectively) indicate the possible photometric ranges for the nucleus due to rotational brightness variations, assuming a peak-to-trough photometric range of $\Delta m = 0.6$~mag, for the best-fit phase functions with $G_R$ as a free parameter and assuming $G_R=0.15$, respectively.  Open circles and dates mark photometry measured from 2017 July to 2017 October where the comet is presumed to be inactive, while blue-filled circles mark photometry obtained in this work where we determine that the comet is likely to be active.
}
\label{figure:phasefunction_358p}
\end{figure}

In order to determine the phase function of 358P's nucleus, we first normalize the measured apparent magnitudes, $m(R,\Delta,\alpha)$ to unit heliocentric and geocentric distances, $R$ and $\Delta$, respectively (i.e., $R=\Delta=1$~AU), where $\alpha$ is the solar phase angle, using
\begin{equation}
m(1,1,\alpha) = m(R,\Delta,\alpha) - 5\log(R\Delta)
\end{equation}
The resulting reduced magnitude, $m(1,1,\alpha)$, remains dependent on the solar phase angle via the solar phase function, as well as the rotational phase of the nucleus at the time of the observations in question.  At this time, the rotational properties of 358P are unknown, and since all of our observations were short-duration ``snap-shot'' observations, we are unable to constrain any of these rotational properties from our available data. Our data also all have relatively low S/N, meaning that any brightness variations within observing sequences cannot be reliably used to construct even partial rotational lightcurves.  Therefore, for phase function fitting, we treat the mean brightness of the object on each night as a single instantaneous photometric point at an arbitrary rotational phase.  Given enough sparsely sampled photometric points, we assume that rotational brightness variations will average to zero, yielding a phase function and absolute magnitude reflecting the object's average brightness over its entire rotational lightcurve.  \changed{For the purposes of determining best-fit phase function parameters, however, we adopt a reasonable lightcurve amplitude of $A=0.30$~mag (corresponding to a peak-to-trough photometric range of $\Delta m_R=0.60$~mag, and an axis ratio for the body as projected on the sky of $(a/b)_N\sim1.7$; cf.\ Equation~\ref{equation:nucleus_axis_ratio1}) as the photometric uncertainty of each data point due to the object's unknown rotational phase at the time of observation, and add this in quadrature with measured photometric uncertainty to derive the final uncertainty used in performing our fitting analysis.}


Omitting the two photometric points from 2017 November and 2017 December from our fitting analysis (due to likely activity at those times; cf.\ Section~\ref{subsection:activity_analysis}), we find best-fit parameters of \changed{$H_R=19.5\pm0.2$~mag and $G_R=-0.22\pm0.13$} for 358P's inactive nucleus, using the standard IAU $H,G$ formalism \citep{bowell1989_astphotmodels_ast2}.  In this formalism, $H$ is the absolute magnitude of an object (i.e., the magnitude at $R=\Delta=1$~AU and $\alpha=0^{\circ}$) and $G$ is sometimes referred to as the slope parameter.  A newer three-parameter formalism (the $H,G_1,G_2$ system) for phase functions \citep{muinonen2010_threeparamphasefunction} is increasingly used in the community in addition to or instead of the $H,G$ system, but in this case, we lack photometric data of sufficient precision over a sufficient phase angle range to achieve a meaningful fit to this more complex function.

We plot our best-fit $H,G$ phase function and the data used to fit it in Figure~\ref{figure:phasefunction_358p}.  As can be seen in the plot, much of our data is clustered in phase angle space (near $\alpha\sim15^{\circ}$).  This clustering means that the resulting values of the slope parameter and, in turn, the absolute magnitude of the best-fit phase function are strongly dependent on individual photometric points at $\alpha=10.0^{\circ}$ and $\alpha=22.3^{\circ}$ that were obtained at unknown rotational phases.

\changed{For reference,} we also plot in Figure~\ref{figure:phasefunction_358p} the best-fit $H,G$ function, for which we find \changed{$H_R=20.01\pm0.06$~mag}, assuming the commonly assumed default value of $G=0.15$ for small solar system bodies when a measured $G$ value is not otherwise available.  We see from the plot that this phase function is still formally consistent with all of the relevant photometric data given the range of possible brightness variations relative to the object's mean brightness due to rotation that we assume in this analysis.  For reference, we also compute best-fit values for a linear phase function for our data, finding an absolute magnitude of $m_R(1,1,0)=19.9\pm0.2$~mag and a phase-darkening coefficient of \changed{$\beta = 0.061\pm0.015$~mag~deg$^{-1}$}.

The effective nucleus radius (in km), $r_N$, of an object with an absolute magnitude of $H_R$ is given by
\begin{equation}
p_Rr_N^2 = (2.24\times10^{16})\times10^{0.4[m_{\odot}-H_R]}
\end{equation}
where $p_R$ is the object's geometric $R$-band albedo, and $m_{\odot}=-27.07$~mag is the absolute $R$-band magnitude of the Sun \citep{hardorp1980_sun2,hartmann1982_cometcolorimetry,hartmann1990_chiron}.  Assuming a geometric $R$-band albedo of $p_R=0.05$, similar to that measured for other MBCs \citep{hsieh2009_albedos}, we estimate an effective nucleus radius for 358P of \changed{$r_N=0.32\pm0.03$~km}.

\subsection{Activity Analysis}\label{subsection:activity_analysis}

\setlength{\tabcolsep}{4.5pt}
\setlength{\extrarowheight}{0em}
\begin{table*}[htb!]
\renewcommand{\thetable}{\arabic{table}}
\centering
\caption{\changed{Analysis of 2012-2013 and 2017 Photometry}}
\smallskip
\begin{tabular}{lrrrD@{$\pm$}DD@{$\pm$}DD@{$\pm$}D}
\tablewidth{0pt}
\hline\hline
\multicolumn{1}{c}{UT Date}
 & \multicolumn{1}{c}{$\nu$$^a$}
 & \multicolumn{1}{c}{$m_{R,t}$$^b$}
 & \multicolumn{1}{c}{$H_{R,t}$$^c$}
 & \multicolumn{4}{c}{$A_d$$^d$}
 & \multicolumn{4}{c}{$M_d$$^e$}
 & \multicolumn{4}{c}{$Af\rho$$^f$}
 \\
\hline
\decimals
2012 Oct 06$^g$ &   7.4 & 19.6$\pm$0.1 & 15.5$\pm$0.3 & 130   & 30   & 45   & 10   & 15   & 3   \\
2012 Oct 08     &   8.0 & 19.9$\pm$0.1 & 15.9$\pm$0.3 & 100   & 20   & 32   &  7   & 12   & 3   \\
2012 Oct 12     &   9.1 & 19.6$\pm$0.0 & 15.7$\pm$0.2 & 120   & 20   & 38   &  7   & 13   & 2   \\
2012 Oct 14     &   9.8 & 19.5$\pm$0.1 & 15.6$\pm$0.2 & 120   & 20   & 39   &  7   & 16   & 3   \\
2012 Oct 15     &  10.0 & 19.4$\pm$0.1 & 15.5$\pm$0.2 & 130   & 20   & 43   &  7   & 19   & 3   \\
2012 Oct 15     &  10.0 & 19.4$\pm$0.1 & 15.6$\pm$0.2 & 130   & 20   & 42   &  7   & 16   & 3   \\
2012 Oct 18     &  10.9 & 19.0$\pm$0.0 & 15.3$\pm$0.2 & 160   & 20   & 54   &  8   & 17   & 2   \\
2012 Oct 19     &  11.2 & 19.1$\pm$0.0 & 15.4$\pm$0.2 & 150   & 20   & 49   &  7   & 17   & 2   \\
2012 Oct 22     &  12.0 & 19.0$\pm$0.0 & 15.4$\pm$0.2 & 150   & 20   & 49   &  6   & 15   & 2   \\
2012 Oct 25     &  12.9 & 19.1$\pm$0.0 & 15.5$\pm$0.2 & 130   & 20   & 44   &  5   & 15   & 2   \\
2012 Nov 08     &  17.0 & 19.0$\pm$0.0 & 15.6$\pm$0.1 & 120   & 10   & 40   &  4   & 16   & 1   \\
2012 Nov 09     &  17.2 & 18.7$\pm$0.0 & 15.3$\pm$0.1 & 160   & 10   & 53   &  5   & 15   & 1   \\
2012 Nov 13     &  18.4 & 18.6$\pm$0.0 & 15.1$\pm$0.1 & 190   & 20   & 65   &  6   & 16   & 2   \\
2012 Nov 14     &  18.7 & 18.8$\pm$0.0 & 15.3$\pm$0.1 & 170   & 20   & 55   &  6   & 16   & 2   \\
2012 Nov 22     &  21.0 & 19.1$\pm$0.0 & 15.4$\pm$0.2 & 150   & 20   & 50   &  7   & 18   & 2   \\
2012 Nov 23     &  21.3 & 18.9$\pm$0.0 & 15.2$\pm$0.2 & 180   & 30   & 60   &  8   & 21   & 3   \\
2012 Dec 18     &  28.3 & 19.5$\pm$0.0 & 14.9$\pm$0.3 & 240   & 60   & 80   & 20   & 18   & 4   \\
2012 Dec 19     &  28.6 & 19.5$\pm$0.0 & 14.9$\pm$0.3 & 240   & 60   & 80   & 20   & 18   & 5   \\
2012 Dec 20     &  28.9 & 19.8$\pm$0.0 & 15.2$\pm$0.3 & 180   & 50   & 60   & 20   & 16   & 4   \\
2013 Jan 08     &  34.2 & 20.4$\pm$0.0 & 15.3$\pm$0.4 & 170   & 50   & 60   & 20   & 15   & 5   \\
2013 Feb 04     &  41.4 & 21.4$\pm$0.1 & 15.9$\pm$0.4 & 100   & 30   & 30   & 10   & 11   & 4   \\
2017 Jul 01$^h$ & 284.9 & 24.7$\pm$0.1 & 19.7$\pm$0.3 &  -0.3 &  0.9 & -0.1 &  0.3 &  0.0 & 0.3 \\
2017 Jul 21     & 289.3 & 24.0$\pm$0.1 & 19.5$\pm$0.2 &   0.0 &  0.8 &  0.0 &  0.3 &  0.0 & 0.3 \\
2017 Sep 17     & 302.9 & 24.3$\pm$0.1 & 19.7$\pm$0.3 &  -0.3 &  0.9 & -0.1 &  0.3 &  0.0 & 0.3 \\
2017 Sep 18     & 303.1 & 24.4$\pm$0.2 & 19.7$\pm$0.3 &  -0.5 &  1.0 & -0.2 &  0.3 &  0.0 & 0.3 \\
2017 Sep 22     & 304.1 & 24.2$\pm$0.1 & 19.4$\pm$0.3 &   0.4 &  1.1 &  0.1 &  0.4 &  0.1 & 0.3 \\
2017 Oct 26     & 312.7 & 25.0$\pm$0.4 & 19.6$\pm$0.6 &  -0.1 &  1.7 &  0.0 &  0.6 &  0.0 & 0.3 \\
2017 Nov 18     & 318.8 & 24.2$\pm$0.1 & 18.6$\pm$0.4 &   5   &  3   &  2   &  1   &  0.9 & 0.5 \\
2017 Dec 18     & 327.0 & 23.2$\pm$0.1 & 17.5$\pm$0.4 &  20   &  7   &  7   &  2   &  2.2 & 0.8 \\
\hline
\hline
\multicolumn{14}{l}{$^a$ True anomaly, in degrees.} \\
\multicolumn{14}{l}{$^b$ Equivalent total apparent $R$-band magnitude.} \\
\multicolumn{14}{l}{$^c$ Equivalent total absolute $R$-band magnitude.} \\
\multicolumn{14}{l}{$^d$ Estimated total scattering surface area of visible ejected dust, in $10^5$ m$^2$.} \\
\multicolumn{14}{l}{$^e$ Estimated total dust mass, in $10^6$ kg, assuming $\rho_d=2500$~kg~m$^{-3}$.} \\
\multicolumn{14}{l}{$^f$ $Af\rho$ values, in cm.} \\
\multicolumn{14}{l}{$^g$ All 2012-2013 photometry from \citet{hsieh2013_p2012t1}.} \\
\multicolumn{14}{l}{$^h$ All 2017 photometry from this work.} \\
\end{tabular}
\label{table:activity_358p}
\end{table*}

\begin{figure}[htb!]
\centerline{\includegraphics[width=\columnwidth]{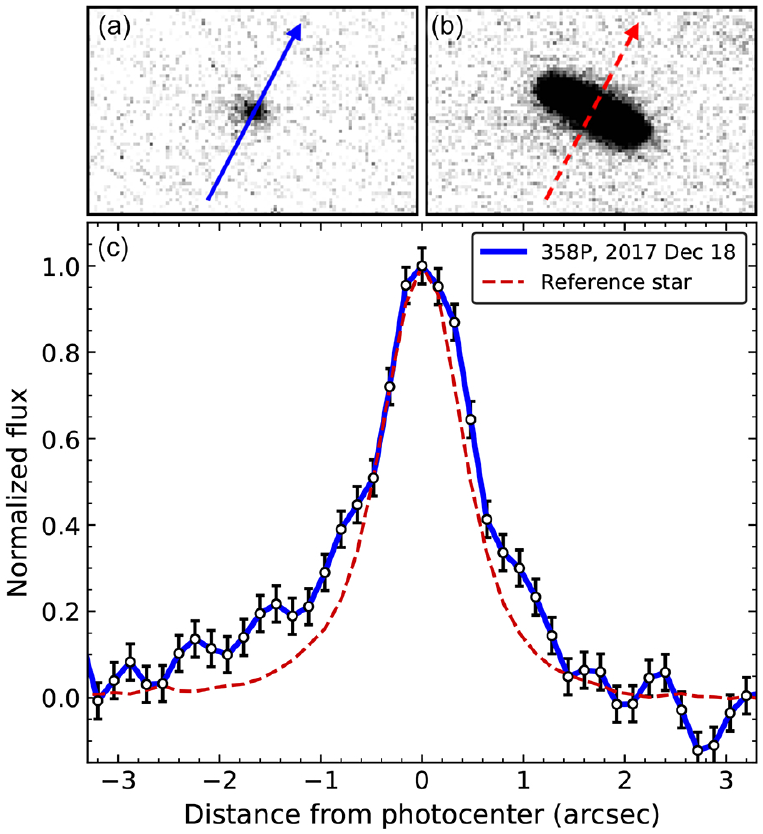}}
\caption{\small (a) Cropped image of the unrotated (i.e., North up, East left) composite image of 358P (center of the panel) from 2017 December 17 showing the position angle and direction of the axis along which the one-dimensional surface brightness profile showing in panel (c) is measured (solid blue arrow). (b) Cropped image of the unrotated composite image of a field star (center of the panel) in our 358P data from 2017 December 17, showing the position angle and direction of the axis along which the one-dimensional surface brightness profile showing in panel (c) is measured (dashed red arrow). (c) One-dimensional normalized surface brightness profile for a composite image of 358P (solid blue line and open circles) on 2017 December 18 overplotted on the one-dimensional normalized surface brightness profile of a corresponding composite image of a reference field star (dashed red line) constructed from the same data set.
}
\label{figure:sbprofile_358p}
\end{figure}

\subsubsection{Activity Detection and Confirmation}\label{subsubsection:activity_detection}

During the course of computing the best-fit phase function of 358P using data acquired in 2017 when the object was apparently inactive (Section~\ref{subsection:phase_function}), we noted that photometric points from 2017 November 18 and 2017 December 18 appeared to be significantly brighter than expected (by $\geq$1~mag) relative to the best-fit function found using only the other data points from 2017 July 1 through 2017 October 26. \changed{A possible faint, short dust tail extending $\sim$1--2 arcsec roughly eastward}, close to the antisolar direction, is also barely visible in the composite image from December (Figure~\ref{figure:images_358p}h).  Careful examination of individual and composite image data did not show any indications of this photometry or observed morphology being contaminated by underlying faint field stars or nearby bright field stars.  The observed photometric deviations are well outside the range of expected rotational variation for the object assuming a physically plausible axis ratio for the nucleus (cf.\ Figure~\ref{figure:phasefunction_358p}).  Measurements of the full width at half-maximum (FWHM) of the point spread functions (PSFs) of the object and nearby field stars did not give any indications of the presence of resolved coma.


To verify the visual detection of activity from 358P on 2017 December 18, we analyze the surface brightness profile of the object on that date.  To do so, we create a composite image constructed by shifting and aligning individual images on the photocenter of a reference star in each frame, where the reference star is chosen based on its proximity to our target object, the fact that it is relatively bright without being saturated, and the absence of other nearby background sources.  This star-aligned composite image and the object-aligned composite image from the same date (Section~\ref{section:observations}; Figure~\ref{figure:images_358p}) are rotated by the same angle such that star trails are horizontal in each image frame.

We then construct one-dimensional surface-brightness profiles oriented along an axis (marked by blue and red arrows superimposed on unrotated composite images of the object and reference star in Figures~\ref{figure:sbprofile_358p}a and \ref{figure:sbprofile_358p}b) perpendicular to the direction of the non-sidereal motion of the object. We measure profiles along this direction because it allows us to directly compare results from images of the object and reference star in a manner that cannot be done along different axes (e.g., the one aligned with the apparent dust tail) due to the trailing of field stars caused by non-sidereal tracking of the telescope to follow the target during the acquisition of these images.

A series of horizontal rectangular apertures are placed along the vertical axes of the rotated images of the object and reference star, where each aperture is one pixel high and the width of the apertures is chosen such that $\sim$90\% of the flux from the source along the horizontal row passing through each source's photocenter is encompassed by the aperture along that central row.  We measure average fluxes within these rectangular apertures, and subtract sky background sampled from nearby areas of blank sky.  

We normalize the profiles of 358P and the reference star to unity at their peaks and plot them together (Figure~\ref{figure:sbprofile_358p}).  Significant excess flux in the direction of the suspected faint dust tail can be clearly seen in the wings of the object profile relative to the stellar profile (while we also see, as noted above, that the FWHMs of the object and stellar profiles are nearly identical).

As can be seen in Figure~\ref{figure:sbprofile_358p}, the requirement that our one-dimensional surface brightness profiles be measured along the axis perpendicular to the non-sidereal velocity vector of the object means that we do not measure the profile along the axis of the apparent dust tail, along which the maximum deviation from a stellar PSF is expected.  As such, the excess flux in the object PSF relative to the stellar PSF shown in Figure~\ref{figure:sbprofile_358p} should be regarded as a lower limit to the true amount of excess flux present in the object's PSF compared to that of an inactive, point-source-like object.

The S/N of the object detection in our 2017 November 18 data is too low to obtain a useful surface brightness profile and perform a similar analysis as described above for our 2017 December 18 data.  As such, we cannot independently corroborate our photometric detection of activity on that date described above.  Nonetheless, based on the photometric indication of activity for our 2017 November 18 data and confirmation of activity via surface brightness profile analysis for our 2017 December 18 data, we conclude that 358P has become active again.  Subtracting the expected brightness of the nucleus from the actual total measured brightness of the object at the time of observations and averaging over a photometry aperture encompassing the visible flux from the object and its apparent dust tail, we estimate the average surface brightness of the tail and any coma that may be present to be $\Sigma_d\sim26.5$~mag~arcsec$^2$.

\subsubsection{Detailed Activity Characterization}\label{subsubsection:dust_mass_analysis}

\begin{figure*}[ht!]
\centerline{\includegraphics[width=5.8in]{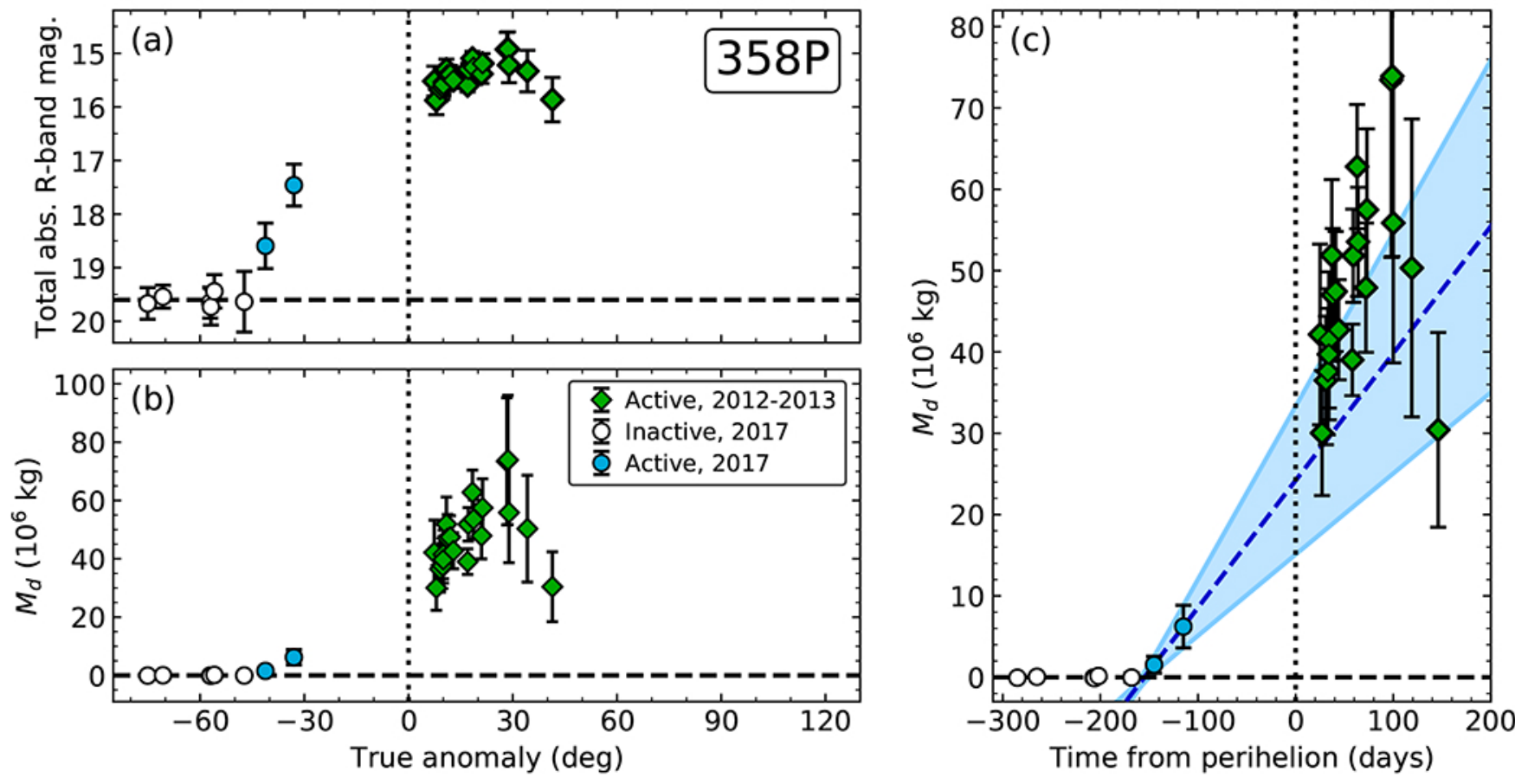}}
\caption{\small (a) Total absolute $R$-band magnitude of 358P during its 2012-2013 active period (green diamonds), 2017 inactive period (open circles), and 2017 active period (blue circles) plotted versus true anomaly ($\nu$).  The expected magnitude of the inactive nucleus is marked with a horizontal dashed black line, while perihelion is marked with a dotted vertical line.  (b) Total estimated dust masses measured for 358P during the same periods of observations as in (a) plotted as a function of $\nu$. The excess dust mass expected for the inactive nucleus (i.e., zero) is marked with a horizontal dashed black line, while perihelion is marked with a dotted vertical line. (c) Estimated total dust masses measured for 358P during 2012-2013 and 2017 plotted versus time from perihelion (where negative values denote time before perihelion and positive values denote time after perihelion). A diagonal dashed blue line shows a linear fit to data obtained on 2017 November 18 and 2017 December 18 ($-41.2^{\circ}<\nu<-33.0^{\circ}$), reflecting an estimate of the average net dust production rate over this period and allowing us to estimate the onset time of activity, while the shaded blue region shows the range of uncertainty of the linear fit.
}
\label{figure:dust_production}
\end{figure*}

Using the phase function derived earlier, we can estimate the amounts of excess dust present in observations from 2012 and 2013 reported by \citet{hsieh2013_p2012t1}, and in 2017 reported in this work.  Following \citet{hsieh2014_324p}, we estimate the total scattering surface area, $A_d$, of visible ejected dust using
\begin{equation}
A_d = \pi r^2_N \left({1-10^{0.4(H_{R,t}-H_R)} \over 10^{0.4(H_{R,t}-H_R)} }\right)
\end{equation}
and the corresponding total mass, $M_d$, using
\begin{equation}
M_d = {4\over 3}\pi r^2_N {\bar a}\rho_d \left({1-10^{0.4(H_{R,t}-H_R)} \over 10^{0.4(H_{R,t}-H_R)} }\right)
\label{equation:dust_mass}
\end{equation}
where $H_{R,tot}$ is the equivalent total absolute magnitude of the active nucleus at $R=\Delta=1$~AU and $\alpha=0^{\circ}$ computed using the $H,G$ phase function and the best-fit $G$ parameter determined above (Section~\ref{subsection:phase_function}; assuming that the dust exhibits the same phase darkening behavior as the nucleus).  We assume dust grain densities of $\rho_d$$\,=\,$2500~kg~m$^{-3}$, consistent with CI and CM carbonaceous chondrites, which are associated with primitive C-type objects like the MBCs \citep{britt2002_astdensities_ast3}.

\citet{moreno2013_p2012t1} found dust grain radii for 358P's observed dust emission in 2012 ranging from $a_{d,{\rm min}}=0.5$~$\mu$m (although they report that this lower limit is not well-constrained) to $a_{d,{\rm max}}=1$$-$10~cm, assuming a power law size distribution with an index of $q=3.5$.  Following \citet{jewitt2014_133p}, we can compute a mean effective particle radius (by mass), ${\bar a_d}$, weighted by the size distribution, scattering cross-section, and residence time, and assuming $a_{d,{\rm max}}\gg a_{d,{\rm min}}$, using
\begin{equation}
{\bar a_d} \sim {a_{d,{\rm max}}\over ln(a_{d,{\rm max}}/a_{d,{\rm min}})}
\end{equation}
Using the particle size distribution determined by \citet{moreno2013_p2012t1}, we obtain ${\bar a_d} = 1$$-$10~mm, where for the purposes of our following analysis, we will use ${\bar a_d} = 1$~mm.

For reference, we also compute $A(\alpha=0^{\circ})f\rho$ values \citep[hereafter, $Af\rho$;][]{ahearn1995_ensemblecomets}, given by
\begin{equation}
Af\rho = {(2R\Delta)^2\over \rho}10^{0.4[m_\odot-m_{R,d}(R,\Delta,0)]}
\end{equation}
where $R$ is in AU, $\Delta$ is in cm, $\rho$ is the physical radius in cm of the photometry aperture used to measure the magnitude of the comet at the distance of the comet, and $m_{R,d}(R,\Delta,0)$ is the phase-angle-corrected (to $\alpha=0^{\circ}$, assuming the same $H,G$ phase function behavior for the dust as for the nucleus) $R$-band magnitude of the excess dust mass of the comet (i.e., with the flux contribution of the nucleus subtracted from the measured total magnitude).  We note, however, that this parameter is not always a reliable measurement of the dust contribution to comet photometry in cases of non-spherically symmetric comae \citep[e.g.,][]{fink2012_afrho}.

The results of all calculations described above are shown in Table~\ref{table:activity_358p} for observations of 358P from \citet{hsieh2013_p2012t1} and this work.  We also plot total absolute magnitudes and computed excess dust masses as functions of true anomaly, $\nu$, and days relative to perihelion in Figure~\ref{figure:dust_production}.

We fit a linear function to the two data points from 2017 when the object appears to be active, aiming to estimate the average dust production rate of the object over this period (represented by the slope of this function) and the start time of its activity.
For reference, the object's heliocentric distance changes from $R=2.522$~AU to $R=2.479$~AU between the two dates at which those data points were acquired.  This heliocentric distance change corresponds to a $\sim$5-20\% increase in the water sublimation rate on the object's surface, depending on whether the isothermal or subsolar approximation is assumed \citep[following the calculations detailed by][]{hsieh2015_ps1mbcs}. The resulting dust production rate and corresponding activity start date we find are of course subject to numerous sources of uncertainty including the non-linearity of the actual dust production rate as a function of heliocentric distance, ordinary photometric calibration uncertainties, uncertainties specifically associated with measuring extended objects (e.g., selection of optimal photometry apertures), and the unknown rotational phases of the object at the times when each photometric point was obtained.

Although the rotational phases of the object at the time of our 2017 November and December observations are unknown, we can estimate the uncertainty that the nucleus's rotational lightcurve imparts on our photometry given the amount of unresolved coma dust (which will tend to damp the amplitude of observed rotational brightness variations) that we have calculated to be present at those times.
The implied minimum axis ratio (neglecting possible projection effects), $(a/b)_N$, for an object with a rotational lightcurve with a measured or assumed peak-to-trough photometric range, $\Delta m_R$, is given by
\begin{equation}
\left({a\over b}\right)_N = 10^{0.4\Delta m_R}
\label{equation:nucleus_axis_ratio1}
\end{equation}
The underlying axis ratio of the nucleus of an active object including coma dust within a given photometry aperture can be computed using
\begin{equation}
\left({a\over b}\right)_N + {F_d\over F_N}\left(1-10^{0.4\Delta m_{\rm obs}}\right) \left({a\over b}\right)^{1/2}_N - 10^{0.4\Delta m_{\rm obs}} = 0
\label{equation:nucleus_axis_ratio}
\end{equation}
where $\Delta m_{\rm obs}$ is the observed peak-to-trough photometric range, and $F_d/F_N$ is the computed dust-to-nucleus flux ratio.  This equation can be solved for $(a/b)^{1/2}_N$ using standard techniques for solving quadratic polynomials \citep{hsieh2011_176p}, where $F_d/F_N=A_d/A_N$ and $A_N=\pi r_N^2=3\times10^5$~m$^2$ for 358P.  Rearranging Equation~\ref{equation:nucleus_axis_ratio}, we can solve for the expected observed photometric range for an object with a known or assumed nucleus axis ratio and measured dust-to-nucleus flux ratio using
\begin{equation}
\Delta m_{\rm obs} = 2.5 \log \left( {\left({a\over b}\right)_N + \left({a\over b}\right)_N^{1/2}\left({F_d\over F_N}\right)}  \over  {1 + \left({a\over b}\right)_N^{1/2}\left({F_d\over F_N}\right)} \right)
\end{equation}
\changed{Assuming the same lightcurve amplitude of $A_N=0.30$~mag (corresponding to $\Delta m_R=0.60$~mag) for 358P's nucleus that we used earlier (Section~\ref{subsection:phase_function})} and the estimated total scattering surface areas of dust in Table~\ref{table:activity_358p}, we then find $\Delta m_{\rm obs}=0.24$~mag (corresponding to an expected observed amplitude of $A_{\rm obs}=0.12$~mag), for the nucleus on 2017 November 18 and $\Delta m_{\rm obs}=0.09$~mag (corresponding to $A_{\rm obs}=0.05$~mag) on 2017 December 18.  We therefore find that for a reasonable assumed axis ratio for 358P's nucleus, lightcurve amplitude damping by coma material should reduce potential rotational brightness variations to comparable or negligible levels relative to measured photometric uncertainties and uncertainties on our best-fit values for $H_R$ and $G_R$ that are used to compute $M_d$ via Equation~\ref{equation:dust_mass}.  We also conclude that any photometric fluctuations observed for 358P in observations obtained in 2012 and 2013, when inferred excess dust masses were much larger than in 2017, are unlikely to be due to nucleus rotation and more likely to be due to observational effects such as seeing fluctuations causing fluctuations in the amount of dust contained within fixed photometry apertures from one image to the next.  Given the slow dust ejection speeds found for most MBCs \citep[e.g.,][]{hsieh2004_133p,hsieh2009_238p,hsieh2011_176p,moreno2011_324p,moreno2016_p2015x6}, we do not expect rotational variations in dust production rates to cause significant fluctuations in measured photometry from ground-based data.

Using the uncertainties we originally calculated for the dust masses measured for 2017 November 18 and 2017 December 18, we estimate an average dust production rate shortly after 358P becomes active of \changed{$\dot M = 2.0\pm0.6$~kg~s$^{-1}$} \changed{\citep[roughly consistent with the average dust production rate found for the object in 2012 by][]{moreno2013_p2012t1}} and an estimated start date of $155\pm4$~days prior to perihelion.  This start date corresponds to 2017 November $8\pm4$ when the object was at $\nu=316.1^{\circ}\pm0.9^{\circ}$, or equivalently, $\nu=-43.9^{\circ}\pm0.9^{\circ}$, and at $R=2.538$~AU.  

\changed{For reference, we perform a simple dust modeling analysis to determine whether the start date we estimate from the object's dust mass evolution is consistent with the activity that we visually detect in our 2017 December 18 observations.  Using an online tool\footnote{\tt http://www.comet-toolbox.com/FP.html} developed by J.~B.\ Vincent for plotting syndyne and synchrone grids \citep{finson1968_cometdustmodeling1}, we find that $a_d\sim1$~mm dust grains ejected with zero initial velocity and evolving under the influence of solar gravity and solar radiation pressure would be expected to travel $\sim$1~arcsec from the nucleus in 40 days (the length of time elapsed between our estimate for the start date of the activity and the observations in question).  Slightly smaller dust grains ($a_d\sim100$~$\mu$m) would be expected to travel about 7~arcsec from the nucleus over the same period of time.  Given that the dust ejected by 358P likely contains a range of dust particle sizes, and not just the mean particle size of ${\bar a_d} = 1$~mm that we use above for our dust mass calculations, we find that these dust modeling results are fully consistent with our observations of a $\sim$1-2~arcsec dust tail on 2017 December 18 and our estimated start date of 2017 November $8\pm4$ for the current active period.  More detailed dust modeling is not justified at this time due to the minimal amount of data presently available for 358P's current active period, although it is planned in the upcoming year once more data can be obtained during the object's next available observing window (see below).
}

Observations when the object is active in 2012-2013 and 2017 do not overlap in true anomaly, and so it is not possible at this time to compare activity levels from the different active epochs to ascertain how much activity attenuation, if any, has taken place.  From Figure~\ref{figure:dust_production}, it appears possible that if the initial dust production rate estimated for 358P in 2017 is assumed (likely incorrectly) to be maintained at a constant level at later times, a similar amount of total dust (within uncertainties) could be ejected by the object during its current active period as was measured during its 2012-2013 active period.  Detailed dust modeling or comparison of observations from both epochs covering overlapping orbit arcs (or ideally both) are needed, however, to robustly compare the strength of the activity observed during each epoch.  Both of these tasks will be the focus of future work, subject to the acquisition of new observations during 358P's next upcoming observing window.

358P is next observable from 2018 August until 2019 May, during which it will cover a true anomaly range of $30^{\circ}\lesssim\nu\lesssim100^{\circ}$, where observability will be best at Northern Hemisphere sites.  This observing window should allow for the acquisition of observations that will overlap the latter portion of the orbital arc covered by previously reported 2012-2013 data and that should also help constrain the rate of fading of residual activity from the object as it approaches aphelion.  Observations of the object are highly encouraged during this period.

\section{Discussion}\label{section:discussion}

\changed{The effective nucleus radius found here for 358P ($r_N=0.32\pm0.03$~km) places it among the smallest MBC nuclei measured to date, along with 238P \citep[$r_N\approx0.4$~km;][]{hsieh2011_238p} and 259P \citep[$r_N=0.30\pm0.02$~km;][]{maclennan2012_259p}.  For comparison, the largest measured MBC nuclei have effective radii of $r_N\sim2$~km \citep[133P and 176P;][]{hsieh2009_albedos}.  Meanwhile, the initial dust production rate estimated here for 358P is generally comparable (within an order of magnitude) to other dust production rates computed for other MBCs \citep[e.g.,][]{hsieh2009_238p,moreno2011_324p,moreno2016_p2015x6,moreno2017_p2016j1,licandro2013_288p,jewitt2014_133p,jewitt2015_313p2}, though we note that these estimated production rates can vary somewhat (within a factor of a few) even for the same active episode for the same object depending on assumptions made about grain densities and mean grain sizes.  We thus find 358P to be similar both in terms of nucleus size and dust production rate with at least some previously characterized MBCs.}

Observations of the reactivation of 358P in 2017 reported in this work make the object the seventh active asteroid in the main asteroid belt to be confirmed to exhibit recurrent activity (i.e., whose activity is very likely to be sublimation-driven, making the object a MBC), after 133P/Elst-Pizarro, 238P/Read, 259P/Garradd, 288P/(2006) VW$_{139}$, 313P/Gibbs, and 324P/La Sagra \citep{hsieh2004_133p,hsieh2011_238p,hsieh2015_313p,hsieh2015_324p,hsieh2017_259p,agarwal2016_288p_cbet}.  The identification of these objects as likely ice-bearing bodies is significant as it increases the number of objects that can potentially be used to help constrain the distribution of icy material in the inner solar system.

In addition to the activity turn-on point determined here for 358P ($\nu\sim315^{\circ}$; $R\sim2.55$~AU), well-constrained turn-on points (where observations of inactivity are followed after a relatively short period of time, e.g., a few months or less, by observations of activity) are now available for a handful of MBCs, including 133P \citep[$\nu\sim350^{\circ}$; $R\sim2.65$~AU;][]{hsieh2010_133p} and 238P \citep[$\nu\sim305^{\circ}$; $R\sim2.60$~AU;][]{hsieh2011_238p}, where 324P has also been detected to be active as early as $\nu\sim300^{\circ}$ \citep[$R\sim2.80$~AU;][]{hsieh2015_324p}.  Compared to these objects, the starting point of 358P's activity in 2017 is unremarkable in terms of $\nu$ or $R$.  As more well-constrained turn-on points for MBCs are determined, it will be useful to search for correlations between true anomalies and heliocentric distances of turn-on points with measured dust production rates and other metrics related to activity strength, and to combine these analyses with thermal modeling to see if these properties can be used to place constraints on other parameters of interest, such as ice depth and quantity \citep[e.g.,][]{schorghofer2016_asteroidice,schorghofer2018_asteroidiceloss}.

The observation of recurrent activity also offers the opportunity to monitor the evolution of activity strength for an individual object from orbit to orbit, provided that data covering overlapping orbit arcs from different active epochs can be acquired, allowing direct comparison of morphology, total scattering surface area of dust, and other indicators of activity strength, or sufficient data during each active epoch can be acquired to allow for average or peak dust production rates to be computed using numerical dust modeling, even if orbit arcs covered by the data in question are not overlapping.  Real-world characterization of the attenuation of MBC activity over time would help to evaluate the applicability of theoretical models of mantle growth and activity attenuation on MBCs \citep[e.g.,][]{kossacki2012_259p} and also provide additional constraints on estimated activation rate calculations based on analyses of detection rates of MBCs found in surveys \citep[e.g.,][]{hsieh2009_htp}.  This type of analysis is beyond the scope of the current work, but does represent an area of growing potential as the amount of data available for characterizing multiple active epochs for individual MBCs increases.

\section{SUMMARY}\label{section:summary}
In this work, we present the following key findings:
\begin{enumerate}
\item{A photometric analysis of observations obtained in 2017 of 358P using Gemini-S while the object appeared inactive yields best-fit IAU phase function parameters of \changed{$H_R=19.5\pm0.2$~mag and $G_R=-0.22\pm0.13$} for the inactive nucleus, corresponding to an effective radius of \changed{$r_N=0.32\pm0.03$~km} (assuming a $R$-band albedo of $p_R=0.05$).}
\item{In the course of our photometric analysis to determine best-fit phase function parameters for the nucleus of 358P, we find that the object appeared significantly brighter than expected in 2017 November and 2017 December given the phase function of the object computed from data obtained prior to that time.  A faint dust tail \changed{extending $\sim$1--2 arcsec roughly eastward}, approximately in the antisolar direction, is also observed in data obtained on 2017 December 18.  The presence of this dust tail is confirmed by surface brightness profile analysis, where we estimate an average surface brightness of the tail and any coma that may be present to be $\Sigma_d\sim26.5$~mag~arcsec$^2$.  We conclude that 358P has become active again for the first time since its previously observed active period in 2012-2013, making it now the seventh MBC candidate confirmed to exhibit recurrent activity with at least one intervening period of inactivity, a strong indication that its activity is sublimation-driven.}
\item{Fitting a linear function to the ejected dust masses inferred from the excess fluxes measured for 358P in 2017 when it is apparently active, we find an estimated average net dust production rate of \changed{$\dot M = 2.0\pm0.6$~kg~s$^{-1}$} and an estimated activity start date of 2017 November $8\pm4$ when the object was at $\nu=316^{\circ}\pm1^{\circ}$ and at $R=2.54$~AU.}
\item{Insufficient data are available at the present time to ascertain whether the strength of 358P's activity has changed between its 2012-2013 and 2017 active periods.  Further observations are highly encouraged during 358P's upcoming observing window (2018 August through 2019 May), during which the orbit arc covered by the object will partially overlap the arc covered by the object during its 2012-2013 apparition, allowing for direct comparison of the strength of the activity observed during the two active epochs.}
\end{enumerate}

\acknowledgments
HHH, MMK, NAM, and SSS acknowledge support from the NASA Solar System Observations program (Grant NNX16AD68G).  We are grateful to J.\ Chavez, J.\ Fuentes, G.\ Gimeno, M.\ Gomez, A.\ Lopez, L.\ Magill, S.\ Margheim, P.\ Prado, M.\ Schwamb, A.\ Shugart, and E.\ Wenderoth for assistance in obtaining observations, and to an anonymous reviewer who helped to improve this manuscript.  This research made use of Astropy, a community-developed core Python package for astronomy, {\tt uncertainties} (version 3.0.2), a python package for calculations with uncertainties by Eric O.\ Lebigot, and scientific software at www.comet-toolbox.com \citep{vincent2014_comettoolbox}. This work also benefited from support by the International Space Science Institute in Bern, Switzerland, through the hosting and provision of financial support for an international team, which was led by C.\ Snodgrass and included HHH and MMK, to discuss the science of MBCs.  This work is based on observations obtained at the Gemini Observatory, which is operated by the Association of Universities for Research in Astronomy, Inc., under a cooperative agreement with the NSF on behalf of the Gemini partnership: the National Science Foundation (United States), the National Research Council (Canada), CONICYT (Chile), Ministerio de Ciencia, Tecnolog\'{i}a e Innovaci\'{o}n Productiva (Argentina), and Minist\'{e}rio da Ci\^{e}ncia, Tecnologia e Inova\c{c}\~{a}o (Brazil).

\vspace{5mm}
\facilities{Gemini South (GMOS-S)}
\software{Astropy \citep{astropy2018_astropy}, IRAF \citep{tody1986_iraf,tody1993_iraf}, {\tt L.A.Cosmic} \citep[][written for python by M.\ Tewes]{vandokkum2001_lacosmic}, {\tt uncertainties} (E.\ O.\ Lebigot), Comet-Toolbox \citep{vincent2014_comettoolbox}}
\bibliographystyle{aasjournal}
\bibliography{hhsieh_refs}   




\end{document}